\documentclass[letterpaper, paper,11pt]{AAS}	

\usepackage{bm}
\usepackage{amsmath}
\usepackage{amssymb}
\usepackage{xcolor}

\usepackage{subfigure}

\usepackage[colorlinks=true, pdfstartview=FitV, linkcolor=black, citecolor= black, urlcolor= black]{hyperref}
\usepackage{overcite}
\usepackage{footnpag}

\usepackage{graphicx}
\usepackage{caption}
\usepackage{subcaption}
\usepackage{float}
\usepackage{comment}

\usepackage[capitalize]{cleveref}

\newtheorem{problem}{Problem}

\Crefname{problem}{Problem}{Problem}
\crefname{example}{Example}{Example}

\newcommand{\setP}{\mathcal{P}}

\newcommand{\vecz}{\mathbf{z}}

\newcommand{\vecu}{\mathbf{u}}

\newcommand{\vecf}{\mathbf{f}}

\newcommand{\vecs}{\mathbf{s}}

\newcommand{\vecp}{\mathbf{p}}

\newcommand{\costH}{\mathcal{H}}
\newcommand{\poi}{\mathrm{POI}}

\PaperNumber{23-334}
\begin{document}

\title{Global Task-Aware Fault Detection Identification For On Orbit Multi-Spacecraft Collaborative Inspection}

\author{Akshita Gupta\thanks{Ph.D. Student, Department of Industrial Engineering, Purdue University.},  
\ Yashwanth Kumar Nakka\thanks{JPL Postdoctoral Scholar, Maritime and Multi-Agent Autonomy group, Jet Propulsion Laboratory, California Institute of Technology},
\ Changrak Choi\thanks{Robotics Technologist, Jet Propulsion Laboratory, California Institute of Technology, Pasadena, CA, 91104, USA}, and
\ Amir Rahmani \thanks{Supervisor, Maritime and Multi-Agent Autonomy group, Jet Propulsion Laboratory, California Institute of Technology, Pasadena, CA, 91104, USA},
\footnote{Authors Akshita and Yashwanth controbuted equally to this work.}
}
\maketitle{} 		

\begin{abstract}
In this paper, we present a global-to-local task-aware fault detection and identification algorithm to detect failures in a multi-spacecraft system performing a collaborative inspection (referred to as global) task. The inspection task is encoded as a cost functional $\costH$ that informs global (task allocation and assignment) and local (agent-level) decision-making. The metric $\costH$ is a function of the inspection sensor model, and the agent full-pose. We use the cost functional $\costH$ to design a metric that compares the expected and actual performance to detect the faulty agent using a threshold.We use higher-order cost gradients $\costH$ to derive a new metric to identify the type of fault, including task-specific sensor fault, an agent-level actuator, and sensor faults. Furthermore, we propose an approach to design adaptive thresholds for each fault mentioned above to incorporate the time dependence of the inspection task. We demonstrate the efficacy of the proposed method emperically, by simulating and detecting faults (such as inspection sensor faults, actuators, and sensor faults) in a low-Earth orbit collaborative spacecraft inspection task using the metrics and the threshold designed using the global task cost $\costH$.
\end{abstract}

\section{Introduction}
Spacecraft swarms enable a new class of flexible and adaptive missions including collaborative inspection~\cite{nakka2021informationConf,nakka2022informationJour}, and in-orbit construction~\cite{foust2020autonomous}. Progress in miniaturization and CubeSat technology~\cite{yost2021state} has further boosted the interest in a technology demonstration of distributed measurement on missions like MarCo~\cite{schoolcraft2017marco} and GRACE~\cite{flechtner2014status}. The multi-spacecraft mission approach offers intrinsic robustness to faults through reconfiguration and improved science data returns through cooperation and collaboration. While multi-spacecraft missions have many advantages, they require complex communication architectures, leading to potential fault propagation. For example, in a leader-follower configuration, any large disturbance in the motion trajectory of one of the followers propagates through the network, leading to network shape deformation. Another example is in a distributed sensing application; if one of the agents behaves in an adversarial manner, the neighboring agents follow the misbehavior due to the underlying consensus framework.

\begin{figure}
    \centering
    \includegraphics[width=0.9\textwidth]{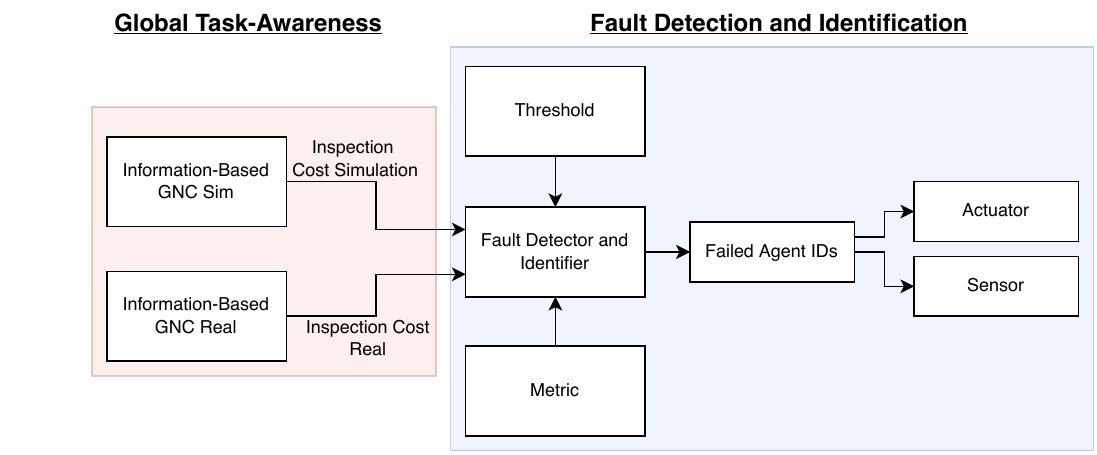}
    \caption{Global task-aware fault detection and isolation for a distributed spacecraft netwrok.}
    \label{fig:fdi-architecture}
\end{figure}

A fault detection, isolation, and recovery (FDIR) architecture is essential to accommodate potential faults at both the network and individual agent levels to continue the mission with graceful degradation. In this work, we present a new FDI method, as shown in~\cref{fig:fdi-architecture}, that detects the failure at the network level using an abstraction of the global task objective $\costH$ and local sensing information and informs the agent level FDIR algorithm to perform necessary actions for recovery. For example, a fault at the network level could be due to communication loss, a global task sensor fault (for inspection task), a global task actuator fault (for on-orbit construction), and a fault at the agent level could be due to thruster or reaction wheels. We propose a simulation vs. real comparison using the $\costH$ and its higher-order gradients. We detect the fault by computing the off-nominal behavior from the expected global task objective $\costH$ by monitoring the individual task using a residual vector sensitive to the agent's faults. The global task objective $\costH$ is designed to be a function of the state of agents in the network and a model of the task sensor (for inspection) or actuator (for construction). The residual vector is a function of the local relative state estimates of the agents in the network. We propose a metric that computes the deviation of the $\costH$ from the expected performance, which is used as an indicator for faults at the network level, and uses higher-order derivatives of $\costH$ to infer if the agent-level faults, as shown in~\cref{fig:fdi-architecture}.

Earlier work~\cite{arrichi2015_IEEE_TCST,paneque2018robust} includes FDI architectures for distributed sensor networks using a local decentralized observer to detect internal agent sensors or actuator failure. Recent work \cite{panagi2011distributed,ramachandran2021resilience} uses adaptive or reconfiguration control with minimal notion of network task to achieve fault tolerance. We instead focus on incorporating global task objectives to inform the local FDIR to react or respond appropriately to continue the mission autonomous to the best possible capability of the network while maximizing the global task objective $\costH$. Furthermore, while most of the earlier work~\cite{boem2018distributed,boem2016distributed} on FDI for distributed systems focuses on simple linear dynamical systems, we focus on low-earth orbit formation flying dynamics that includes periodic orbits.

The main contributions of this work are as follows: 1) we propose an architecture for FDI in a multi-agent spacecraft system that integrates global-task objectives and local-agent level behaviors for task awareness, 2) we derive a global cost functional that is decomposable to cost functions that inform local progress and intermediate consensus on global progress, 3) we propose a novel FDI metric based on the global cost $\costH$ and the high-order derivatives of the $\costH$ to detect and identify both the global and local faults. 

We apply our FDI architecture to a recently proposed multi-agent collaborative spacecraft inspection mission~\cite{nakka2021informationConf, nakka2022informationJour} in a low Earth orbit to detect failures in the inspection sensors and individual agent sensing. The cost function $\costH$ defines the global inspection progress by fusing individual agent sensor data measurements. The inspection data fusion runs at a fixed frequency $\omega_g$, and the network fault diagnosis is run at frequency $\omega_{\mathrm{FDI}}$. We assume that agents communicate with each other only when within the communication radius, leading to a time-varying communication topology and sensing graph. The proposed method is capable of handling the time-varying graph and intermittent communication. We demonstrate that the proposed menthod can detect and identify the faults while keeping track of the global task. This approach is essential to inform the recovery procedure, described in our recent work~\cite{Choi2023Resilient}, for designing new orbits and pointing trajectories to complete the mission.


The paper is divided into four sections: 1) overview of the GNC architecture for the multi-agent collaborative inspection; 2) problem description and an outline of the potential faults at both the global task level and agent level; 3) discussion of the proposed approach and the derivation of the metrics and the threshold for fault detection and identification; and 4) preliminary implementation and simulation results.

\section{Overview of GNC Architecture for Inspection}

In this section, we give a brief overview of the collaborative low-earth orbit inspection framework proposed in our earlier work~\cite{nakka2021information,nakka2022informationJour}. Using this framework we design optimal Passive Relative Orbits\cite{alfriend2009spacecraft,nakka2022informationJour} (PROs) and attitude trajectories for $N$ observer spacecrafts, inspecting $M$ Points of Interest (POIs) on a target spacecraft, by solving the following information-based optimal control problem.

\begin{problem}
Information-Based Optimal Control Problem
\label{prob:info_multi_agent_control}
\begin{align}
    & \underset{\vecp,\vecu_i}{\min} \quad \int_{0}^{t_f} \left(\sum_{j=1}^M \costH (\vecp,\vecs_j) + \sum_{i=1}^N \|\vecu_i\| \right) dt\label{eq:info_problem_cost}\\
    \textbf{s.t.} & \quad \begin{cases}
    \mathrm{Dynamics \ Model:} \quad &\dot \vecp_i = \vecf(\vecp_i,\vecu_i)\\
    \mathrm{Safe \ Set:} \quad &\vecp_i \in \setP, \ \forall i \in \{1,\dots,N\}\\
    \mathrm{Inspection \ Sensor \ Model:} \quad & \vecz_{i,j} = h(\vecp_i,\vecs_j) + \xi, \ \xi \sim \mathcal{N}\left(0,\Sigma_h(\vecp_i,\vecs_j)\right),
    \end{cases} \\
 & \mathrm{Points\ of\ Interest:} \quad \vecs_j \ \forall j \in \{1,\dots, M\} \label{eq:info_problem_constraints}
\end{align}
\end{problem} where $\sum_{j} \costH(\vecp,\vecs_j)$ is the information cost, $\sum_{i} \|\vecu_i\|$ is the fuel cost, $\vecp_i$ is the full-pose of the observer spacecraft, $\vecs_j$ is the full-pose of the $j^{\mathrm{th}}$ $\poi$ on the target spacecraft. The inspection sensor model in~\cref{eq:info_problem_constraints} outputs the value of interest $\vecz_{i,j}$, when the $i^{\mathrm{th}}$ observer with pose $\vecp_i$ is inspecting a $\poi$ at $\vecs_{j}$. Minimizing the information cost $\sum_{j} \costH(\vecp,\vecs_j)$ ensures that the inspection task is complete.

We decompose the~\cref{prob:info_multi_agent_control} to derive a hierarchical GNC algorithm (for details refer to our earlier work~\cite{nakka2021information}). The hierarchical algorithm uses the information-cost and the sensor model to select the informative PROs and attitude pointing vector for each agent. We optimize the informative PROs and attitude plan for optimal orbit insertion, reconfiguration, and attitude tracking using an optimal control problem formulation that computes minimum fuel trajectory using sequential convex programming approach. In this work, we use the information cost to keep track of the task progress and detect off-nominal behaviour of the multi-agent system and the individual agents. We describe the cost functional used to compute the information gain in the following.

\subsubsection{Information Gain.}
\label{subsubsec:information_gain}
To quantify the information, a prior model of the target spacecraft is used along with sampled \emph{points of interest} (POIs) on the surface of the spacecraft. The cost function $\costH$ is designed to minimize the total variance on the knowledge of POIs. We use the cost function $\costH$ designed in~\cite{DBLP:journals/ijrr/SchwagerRS11}, and is a function of POIs as given below.
\begin{align}
    \costH_{\poi}(\vecs) = \left( w^{-1} + \sum_{\vecp\in\setP} \sigma(\vecp,\vecs)^{-1} \right)^{-1}\nonumber\\
    \costH = \sum_{\vecs\in\text{POIs}} \costH_{\poi}(\vecs) \phi(\vecs),
    \label{eq:information_gain}
\end{align}
where $\vecs \in \mathbb{R}^3$ is a POI on the target spacecraft's surface, $w \in \mathbb{R}$ is the initial variance based on the prior model of the target spacecraft, $\vecp \in SE(3)$ is the pose of a sensor mounted on a spacecraft such as a camera, $\setP$ is the set of all sensor poses, $\sigma ( \vecp, \vecs)$ estimates the variance of estimating POI at $\vecs$ with the sensor at $\vecp$, and $\phi(\vecs) \in \mathbb{R}$ is the relative importance of POI $\vecs$. 

The function $\sigma (\cdot,\cdot)$ corresponds to information per pixel. It incorporates sensor chracteristics such as the current uncertainty of the spacecraft's pose estimate, the accuracy of the sensor based on the distance between $\vecp$ and $\vecs$, or the lighting conditions. Here, we use a simple RGB camera sensor and no environmental noise~\cite{DBLP:journals/ijrr/SchwagerRS11}:
\begin{equation}
\sigma (\vecp,\vecs) \propto
\begin{cases}
 \operatorname{dist}^2(\vecp,\vecs) & \vecs \text{ visible from } \vecp\\
\infty & \text{otherwise}
\end{cases},
\label{eq:variance_function}
\end{equation}
where $\operatorname{dist}(\vecp,\vecs)$ is the Euclidean distance between POI $\vecs$ and pose $\vecp$. We compute $\sigma$ using visbility checking. The offline solution to problem \ref{prob:info_multi_agent_control} is used to predict the nominal system behavior in terms of the information-based cost $\costH_{nom}$ over a finite time interval (1 or 2 orbits). As described in~\cref{fig:fdi-architecture}, we precompute the nominal behaviour $\costH_{nom}$ and compare it to the real-time behaviour $\costH$ over the time hotizon $t$ as follows:
\begin{equation}
    \int_{0}^{t}(\boldsymbol{\costH} - \boldsymbol{\costH}_{\mathrm{nom}} ) dt \geq \Delta \costH_{\mathrm{threshold}} t.  
\end{equation}
If the real-time value deviates from the nominal behavior by a threshold $\Delta \costH_{\mathrm{threshold}}$ then a fault is detected.
In the following section, we discuss on how we modify the cost function to construct the FDI architecture in~\cref{fig:fdi-architecture}.
\begin{figure}[h]
    \centering
    \includegraphics[width=0.7\textwidth]{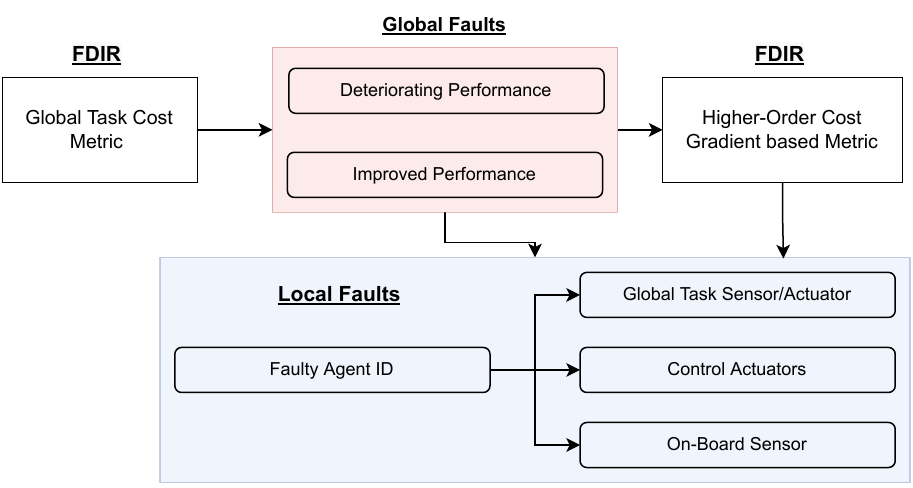}
    \caption{An overviwe of the type of global and local faults detected and identified using the proposed FDI architecture, metrics and the threshold.}
    \label{fig:faults}
\end{figure}

\section{Problem Description}
The goal of the work is to detect the global and local faults in a multi-spacecraft system performing global tasks such as: on-orbit inspection and on-orbit construction. The global behaviour faults and local faults detected using the proposed approach are described in the following Figure~\ref{fig:faults}. Note that the global behaviour fault for inspection task could lead to both deteriorating and improved performance. The improved performance is due two reasons: 1) failure in the controller at the agent level that leads to better exploration of the inpection target, and 2) suprious signals being communicated by the neighbours. The deteriorating performance is due to failure in the inspection sensor or the pointing controller of the spacecraft. The behaviour fault detection is used as an input to identify the agent and the type of fault. The fault detection and identification problem addressed in this work is summarized as follows: 
\begin{problem}\label{prob:main_problem}
    Given the nominal expected global task performance $\costH_{nom}$ and the real-time performance $\costH$, detect the global and agent-level faults in the multi-spacecraft system performing a global task (collaborative inspection or construction). The global and agent-level fault tree is described in the~\cref{fig:faults}.
\end{problem}
In the following sections, we describe the derivation of the metrics and threshold used in the two FDIR blocks shown in Figure~\ref{fig:faults} for fault detection and identification.

\section{Global Task Aware Fault Detection and Isolation}

In this section, we describe the different components of the global task aware fault detection and isolation algorithm that solves the~\cref{prob:main_problem}. This section describes the different components of the fault detection system. We first describe the derivation of the attribute required to detect a faulty spacecraft and the gradient-based fault metric is derived. Finally, we discuss the design of thresholds for detecting and identifying different types of fault.

\subsection{Global Task Cost Functional}
We use a centralized monitering system that utilizes only the information cost updates and implicitly the variance updates of the POIs being shared on the communication network during the collaborative inspection process by the individual spacecraft to detect off-nominal behaviour and identify the faults . At any given time instant $t$, each spacecraft $i$ shares its local information cost value $\costH_i(t)$ with the central computing system. To determine the value of $\costH_i(t)$ observe that from \eqref{eq:information_gain}, the centralized cost function is given as follows:
\begin{equation}
    \begin{split}
        \boldsymbol{\costH} &= \sum_{s \in \mathcal{S}} \costH_{POI}(s) \phi(s),\\
        \text{Where: } \costH_{POI}(s) &= ( w^{-1} + \sum_{p \in \mathcal{P}} f(p,s)^{-1} )^{-1}. 
    \end{split}
\end{equation}
The above cost functional for individual POI can be decomposed into contributions from individual agent $H_i$ and a consensus term $\psi(s)$ that is updated over fixed intervals of time.
\begin{equation}
    \begin{split}
        \costH_{POI}(s) &= ( w^{-1} + \sum_{p \in \mathcal{P}} f(p,s)^{-1} )^{-1} \\
        \costH_{POI}(s) &= \psi(s) (w^{-1} + \sum_{p \in \mathcal{P}} f(p,s)^{-1})
    \end{split}
\end{equation}
where, $\psi(s) = \frac{1}{(w^{-1} + \sum_{p \in \mathcal{P}} f(p,s)^{-1})^2}$. Therefore,
\begin{equation}
    \begin{split}
        \boldsymbol{\costH} &= \sum_{s \in \mathcal{S}} \phi(s) \psi(s) (w^{-1} + \sum_{p \in \mathcal{P}} f(p,s)^{-1}) \\
        \boldsymbol{\costH} &= \sum_{s \in \mathcal{S}} \phi(s) \psi(s) w^{-1} + \underbrace{ \sum_{s \in \mathcal{S}_1} \phi(s) \psi(s) f(p_1, s)^{-1} }_\text{$H_1$} + \dots + \underbrace{ \sum_{s \in \mathcal{S}_N} \phi(s) \psi(s) f(p_n, s)^{-1} }_\text{$H_N$}. \\
        \boldsymbol{\costH} &= \sum_{s \in \mathcal{S}} \phi(s) \psi(s) w^{-1} + \sum_{p_i \in \mathcal{P}} \underbrace{\sum_{s \in \mathcal{S}_i} \phi(s) \psi(s) f(p_i, s)^{-1} }_\text{$H_i$}
    \end{split}
    \label{eqn:decomposed_cost}
\end{equation}

\noindent Using \eqref{eqn:decomposed_cost}, the information cost for each spacecraft $i$ is computed as $\costH_i(t) = \sum_{s \in \mathcal{S}_i} \phi(s) \psi(s) f(p_i(t), s)^{-1}$ where $\psi(s)$ is the normalization factor. This attribute not only tracks the performance of individual spacecrafts, but also prevents the need to explicitly share information about the state of spacecrafts over the communication channel inbetween the agents. 

\noindent \textbf{Note:} The new formulation in~\cref{eqn:decomposed_cost} is a linear combination of the individual contributions by each agent $i$ at high frequency updates and intermediate consensus term $\psi(s)$ that is updated at low frequency. This decomposition, is useful in computing the performance of individual agent, and design the metric and threshold using the individual agent contribution $H_i$.

\subsection{Fault Metric}

In this work, a gradient based metric is used to compare the actual progress made by individual spacecrafts to the expected task progress in a finite time interval $\Delta t$. Let $\costH_i^{pred}(t)$ denote the expected information cost for spacecraft $i$ at time $t$, computed by the central computing system using a simulation setup. Then the fault detection metric for a particular spacecraft is given as follows:
\begin{equation}
    \costH_{m_i}(t) = \textit{abs} \left( 1 - \frac{\Delta \costH_i(t) }{\Delta \costH_i^{pred}(t) } \right) = \textit{abs} \left( 1 - \frac{\costH_i(p_i[t]) - \costH_i(p_i[t-\Delta t])}{\costH_i^{pred}(p_i[t]) - \costH_i(p_i[t-\Delta t])} \right).
\label{eqn:fault_metric}
\end{equation} 
The above metric can be used to distinguish between whether a spacecraft is exhibiting faulty behavior or not as follows:
\begin{equation}
    \costH_{m_i}(t) 
    \begin{cases}
    = 0 & \text{ No fault} \\
    > 0 & \text{ Fault has occured }. \\
    \end{cases}
    \label{eqn:fault_detection_condition}
\end{equation}
Note that the occurence of a fault can also improve the information gain, i.e., a faulty spacecraft can perform better than expected.
However, there is still a need to identify this type of fault because the spacecraft is not behaving as expected. The fault metric in \eqref{eqn:fault_metric} can also be used to determine if the performance of a spacecraft has improved or deteriated. Defining $x := \frac{\Delta \costH_i(t)}{\Delta \costH_i^{pred}(t)}$, table 1 below specifies the criteria to analyze the performance of a faulty spacecraft $i$.

\begin{table}[!h]
\begin{center}
\begin{tabular}{| c | p{80mm} | }
\hline
 \textbf{Fault Case} & \textbf{Condition} \\ 
 \hline
  Deteriorating performance & $\textit{sign} \left( \Delta \costH_i(t) \right) \neq \textit{sign} \left( \Delta \costH_i^{pred}(t) \right)$; \newline
  $\textit{sign} \left( \Delta \costH_i(t) \right) = \textit{sign} \left( \Delta \costH_i^{pred}(t) \right)$ \textbf{and} $x<1$\\
  \hline
  Improved performance & $\textit{sign} \left( \Delta \costH_i(t) \right) = \textit{sign} \left( \Delta \costH_i^{pred}(t) \right)$ \textbf{and} $x>1$\\
  \hline
\end{tabular}
\caption{Identifying spacecraft performance under fault.}
\end{center}
\end{table}
\vspace{-10mm}
\subsection{Fault Type}
The primary goal of the centralized FDIR system is to identify the faulty observer spacecrafts in the network. This work is primarily focused on detecting actuator and sensor faults in the spacecraft. In the GNC architecture, an actuator fault can either occur during the state propagation of a spacecraft or during the optimal sensor placement to observe the target POIs.

In the former case, the faulty spacecraft doesn't follow its assigned orbit, thereby changing the set of POIs which it observes. Such a faulty observer behaves as a rogue spacecraft or can collide with another spacecraft in the vicinity. In the latter case, the onboard sensor fails to accurately point at the POI with maximum variance. Both these fault cases will cause a change in the information gain computed by the observer. In the case of a sensor fault, the variance in observing a POI is different than expected.

\begin{figure}[h]
    \centering
    \includegraphics[width=0.48\textwidth]{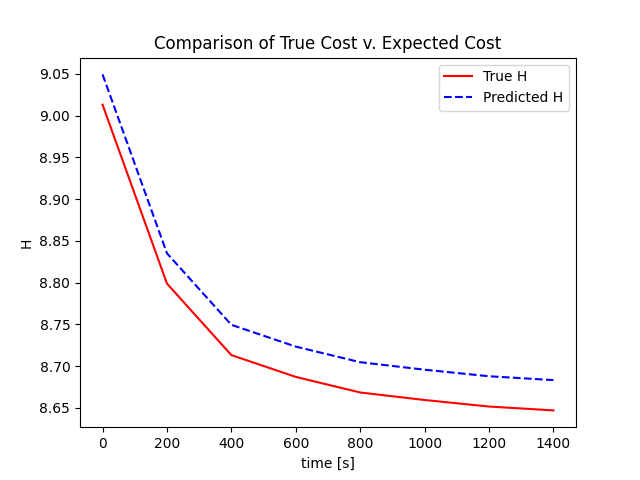}
    \includegraphics[width=0.48\textwidth]{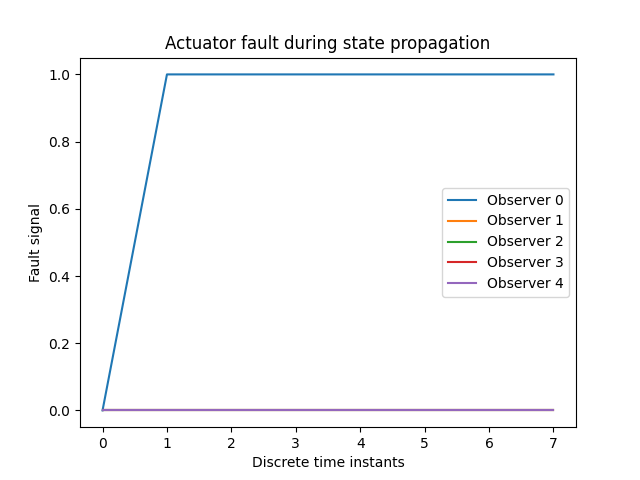}
    \caption{real-time vs. Expected cost under actuator attack I (left); Behavior of fault signal (right).}
    \label{fig:actuator1}
\end{figure}

\begin{figure}[h]
    \centering
    \includegraphics[width=0.48\textwidth]{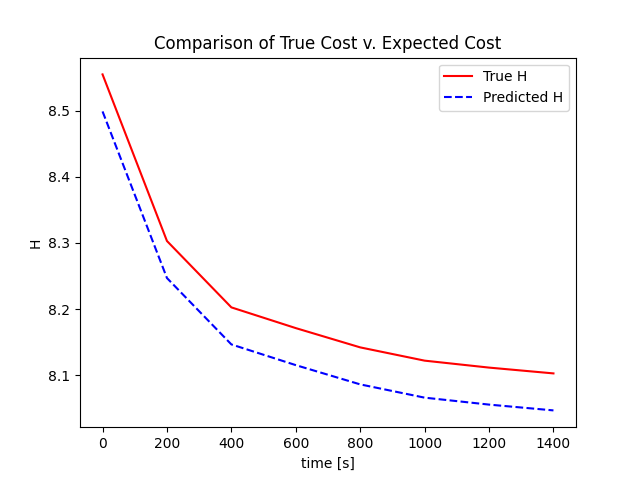}
    \includegraphics[width=0.48\textwidth]{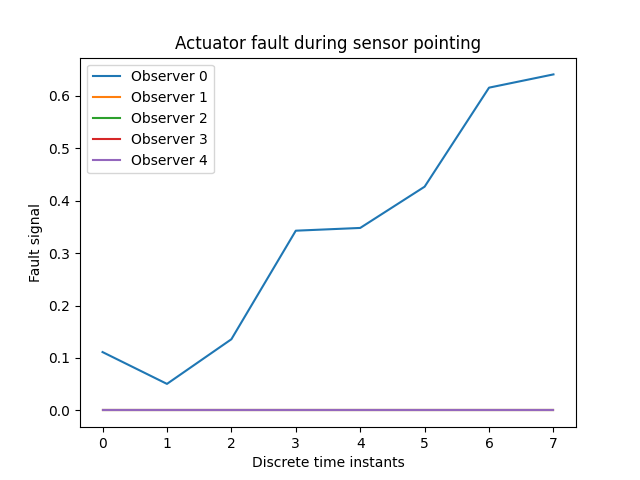}
    \caption{Real-time vs. Expected cost under actuator attack II (left); Behavior of fault signal (right).}
    \label{fig:actuator2}
\end{figure}

Figures \ref{fig:actuator1} and \ref{fig:actuator2} demonstrate the effect of different types of actuator faults on the global information gain $\costH$, as well as the fault signal for different observer spacecrafts. The actuator faults are simulated on the observer spacecraft 0 by adding a random noise to the state of the observer and the pose of the onboard sensor, respectively. In Fig. \ref{fig:actuator1} the occurence of actuator fault improves the global system performance since the true value of information gain is lower than the expected value. The corresponding plot for the fault signal shows a value of 0 for non-faulty observers and a value of 1 for the faulty observer.
This is consitent with the conditions in \eqref{eqn:fault_detection_condition}.
Figure \ref{fig:actuator2} shows a similar behavior when the actuator fault is implemented during sensor pointing.
In this case the global system performance deteriorates.

The above fault cases show that the proposed fault metric performs as expected. However, there is still a need to determine an appropriate threshold to distinguish between system noise and a fault. The occurence of actuator faults cause a change in the pose $p_i(t)$ of a spacecraft. From the formulation of the global cost $\costH$ in \eqref{eqn:decomposed_cost}, it is explicitly clear that a change in pose $p_i(t)$ of a spacecraft will effect the variance in observing a target POI. There is also a more subtle dependence of the set of visible POIs, $\mathcal{S}_i(t)$, on the pose of spacecraft, since the field of view of the onboard sensor will change w.r.t. the spacecraft pose.

Let $\mathcal{S}_i^{pred}(t)$ be the expected set of visible POIs for spacecraft $i$ at time $t$. Then, to compute a fault threshold, it is first necessary to construct a set of POIs $\mathcal{S}_i^{'}(t)$, such that
\begin{equation}
0 < | \costH_i(\mathcal{S}_i^{'}(t)) - \costH_i(\mathcal{S}_i^{pred}(t)) | \le | \costH_i(\mathcal{S}_i(t)) - \costH_i(\mathcal{S}_i^{pred}(t)) |, \quad \quad \forall \mathcal{S}_i(t) \subset \mathcal{S}.
\label{eqn:fault_set}
\end{equation}

The fault threshold for individual spacecrafts can be computed as
\begin{equation}
\tau_i(t) = \textit{abs} \left( 1 - \frac{\costH_i(\mathcal{S}_i^{'}(t)) - \costH_i(\mathcal{S}_i(t - \Delta t)) }{ \costH_i(\mathcal{S}_i^{pred}(t)) - \costH_i(\mathcal{S}_i(t - \Delta t)) }\right).
\label{eqn:fault_threshold}
\end{equation}
Equation \eqref{eqn:fault_threshold} is used to construct adaptive fault thresholds for individual spacecrafts, where an actuator fault is detected in spacecraft $i$ if $\costH_{m_i}(t) > \tau_i(t)$.

\section{Implementation and Results}

\begin{figure}[!h]
    \centering
    \includegraphics[width=0.49\textwidth]{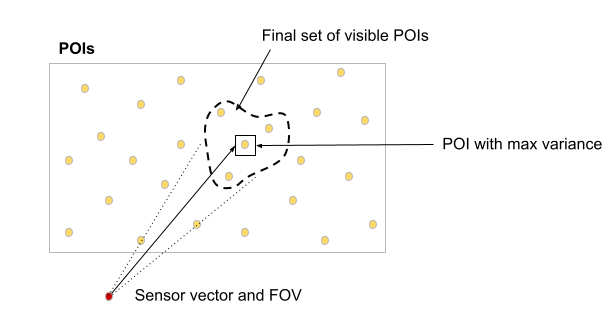}
    \includegraphics[width=0.49\textwidth]{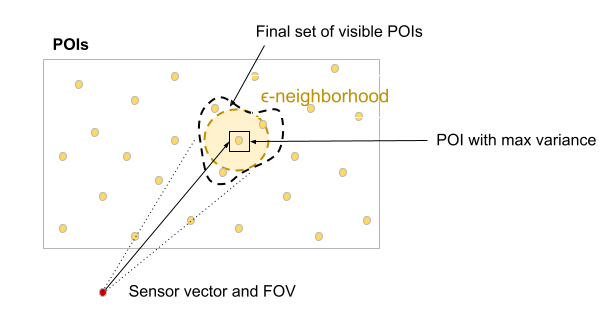}
    \caption{Visible set of POIs for a spacecraft (left); $\epsilon-$neighborhood constructed around POI with maximum variance (right).}
    \label{fig:POI_selection}
\end{figure}

\color{black}
Computing the set $\mathcal{S}_i^{'}(t)$ in \eqref{eqn:fault_set} can become computationally intractable with increasing number of POIs. Therefore, a sampling based approach is taken to approximate the set $\mathcal{S}_i^{'}(t)$ by randomly pointing the onboard sensor within an $\epsilon-$neighborhood of the target POI. Figure \ref{fig:POI_selection} demonstrates the construction of this $\epsilon-$neighborhood, where the onboard sensor vector is randomly pointed to any point in the $\epsilon-$neighborhood, thereby changing its FOV and the visible set of POIs. For actuator faults, the value of $\epsilon$ can be estimated by analyzing $\frac{\partial p_i}{\partial u_i}$ for the system.

The fault detection framework was incorporated in the simulation setup for the hierarchical planning algorithm \cite{nakka2021information, nakka2022informationJour}. During the orbit assignment phase at time $t$, the central agent determines the expected set of visible POIs, $\mathcal{S}_i^{pred}(t)$, for each spacecraft $i$. This is used to compute the nominal behavior, $\costH_i(\mathcal{S}_i^{pred}(t))$, for each spacecraft during a fixed time period of 2 orbits. The fault threshold, $\tau_i(t)$, for each spacecraft is also computed during the orbit assignment phase. In the following plots, 10 target POIs were sampled in an $\epsilon-$neighborhood around the POI with maximum variance, for each spacecraft. The sampled set $\mathcal{S}_i^{'}(t)$ which gives the minimal value for $\tau_i(t)$ determines the fault threshold for spacecraft $i$ at time $t$. At the local agent level, each spacecraft receives its orbit assignment and tracks the progress of its local information cost, $\costH_i(t)$, while propagating the next 2 orbits. At the end of this fixed time interval, each spacecraft transmits its local information cost to the central agent where the centralized FDIR algorithm detects any faulty spacecraft behavior using the metric in \eqref{eqn:fault_metric}.

\color{black}

\begin{figure}[!h]
    \centering
    \includegraphics[width=0.46\textwidth]{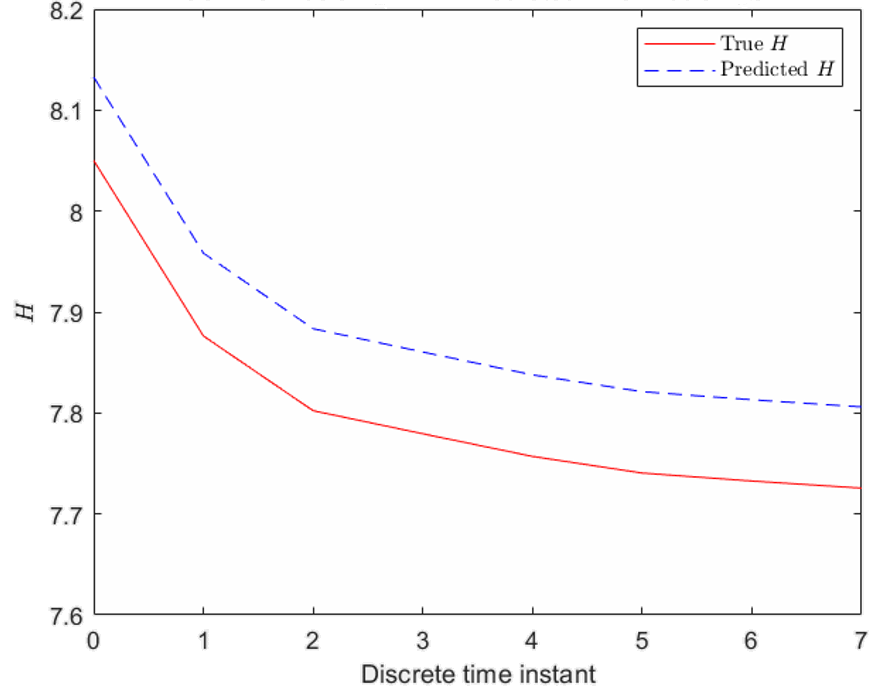}
    \includegraphics[width=0.48\textwidth]{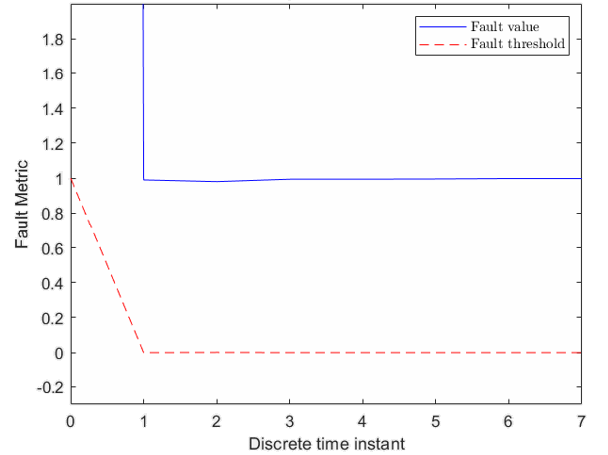}
    \caption{Real-time vs. Expected cost under actuator attack I (left); Adaptive threshold for faulty spacecraft 4 (right).}
    \label{fig:threshold1}
\end{figure}

\begin{figure}[!h]
    \centering
    \includegraphics[width=0.46\textwidth]{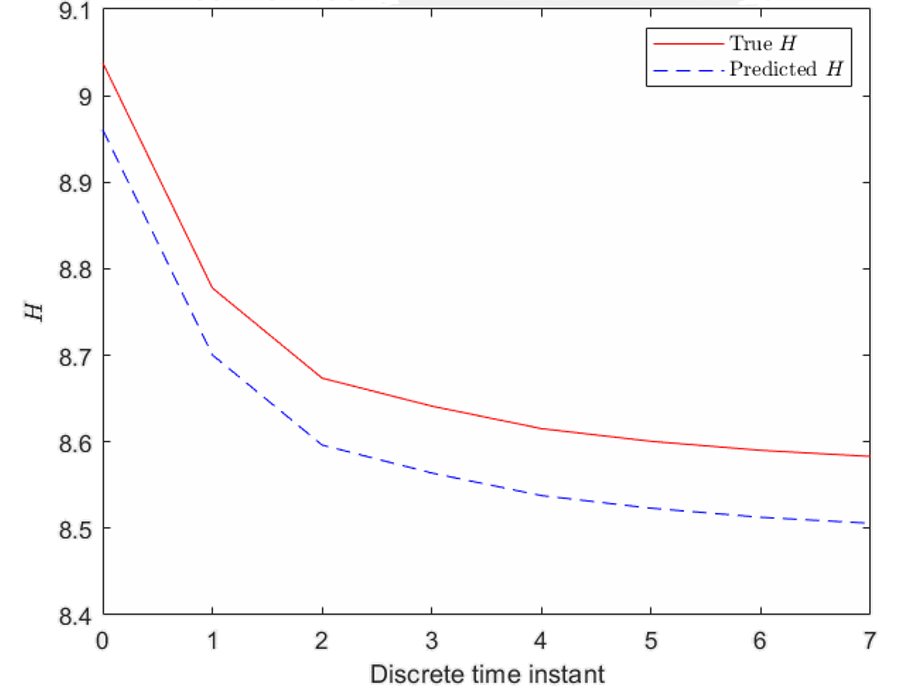}
    \includegraphics[width=0.48\textwidth]{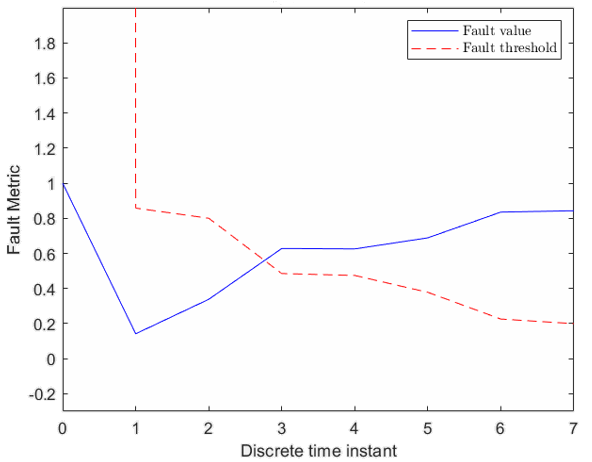}
    \caption{Real-time vs. Expected cost under actuator attack II (left); Adaptive threshold for faulty spacecraft 3 (right).}
    \label{fig:threshold2}
\end{figure}

The performance of the above technique is tested on different types of actuator faults, as shown in Fig. \ref{fig:threshold1} and \ref{fig:threshold2}.
In Fig. \ref{fig:threshold1}, an actuator fault is implemented during the state propagation of spacecraft 4.
In this case, the overall task performance improves and the proposed faul threshold succesfully detects the actuator fault.
In Fig. \ref{fig:threshold2}, an actuator fault is implemented during the sensor pointing phase causing the overall system performance to deteriorate.
The proposed adaptive threshold detects the fault after first few time instants.


It is noted that the sampling based threshold design is susceptible to false negatives if the samples are not uniformly distributed or the number of samples are not sufficient to construct a tight estimate. Finally, designing a threshold for sensor faults depends on the hardware specifications.

\section{Conclusion}
This work presents a global-to-local task-aware fault detection and identification method for a multi-spacecraft system. When coupled with a replanning method, the algorithm provides a framework for graceful degradation in performance while ensuring mission completion. We derived a gradient-based fault metric used by a central computing system to detect faulty spacecraft in the system in a multi-agent collaborative task such as on-orbit inspection and construction. The fault metric is derived using a global task cost functional that encodes the task of the swarm as a decomposable cost functional for individual agents, while retaining the capability of having an intermediate consensus among locally neighboring agents. This metric is implemented on a low-earth on-orbit collaborative inspection to detect performance deviation from faults and identify the fault type. Further, an adaptive fault threshold is designed to identify actuator faults in individual spacecraft. We demonstrate the approach on the inspection mission by detecting faults using only the global cost metric and identifying the fault type using the higher order gradients of the cost metric.

\section*{Acknowledgement}
The research was carried out at the Jet Propulsion Laboratory, California Institute of Technology, under a contract with the National Aeronautics and Space Administration © 2022. California Institute of Technology. Government sponsorship acknowledged. 

\bibliographystyle{AAS_publication}   
\bibliography{main}   
\end{document}